\documentclass[12pt]{iopart}
\usepackage{iopams}

\def\Journal#1#2#3#4{{#4} {\it #1} {\bf #2}, #3 }
\def\a{\alpha}
\def\b{\beta}
\def\g{\gamma}
\def\d{\delta}
\def\e{\epsilon}

\begin{document}

\title{Silent universes with a cosmological constant}

\author{Norbert Van den Bergh and Lode Wylleman~\footnote{research assistant supported by
the Fund for Scientific Research Flanders(F.W.O.), e-mail:
lwyllema@cage.ugent.be}}

\address{Faculty of Applied Sciences TW16, Gent University, Galglaan 2, 9000 Gent, Belgium}

\begin{abstract}
We study non-degenerate (Petrov type I) silent universes in the
presence of a non-vanishing cosmological constant $\Lambda$. In
contrast to the $\Lambda=0$ case, for which the orthogonally
spatially homogeneous Bianchi type I metrics most likely are the
only admissible metrics, solutions are shown to exist when
$\Lambda > 0$. The general solution is presented for the case
where one of the eigenvalues of the expansion tensor is 0.

\end{abstract}

\pacs{0420,0440}



\section{Introduction}

The electric and magnetic part of the Weyl tensor with respect to some unit timelike vector
field $u^a$ are defined by
\begin{equation}\label{EenH}
E_{ab}\equiv C_{acbd}u^c u^d   \quad  H_{ab}\equiv C^*_{acbd}u^c
u^d = \frac{1}{2}{\eta_{ac}}^{mn}C_{mnbd}u^c u^d
\end{equation}
with $\eta_{abcd}$ the 4-dimensional Levi-Civita tensor.
\emph{Silent universes} are defined \cite{Matarrese} as
irrotational expanding dust models, which are solutions of the
Einstein field equations
\begin{equation}\label{veldvergelijkingen}
R_{ab}-\frac{1}{2}R g_{ab}+\Lambda g_{ab}=T_{ab} = \mu u_a u_b
\end{equation}
(wherein $\mu\neq 0$) with the additional property that the Weyl tensor is \emph{purely
electric}  with respect to the fluid 4-velocity $u^a$:
\begin{equation}
H_{ab} = 0.
\end{equation}

The Petrov type is then necessarily O, D or I~\cite{Barnes}, with
the type O solutions given by the Friedman-Robertson-Walker dust
metrics. All Petrov type D solutions are known explicitly too:
they are characterised~\cite{Barnes} by the fact that the Weyl
tensor is degenerate in the same plane as the shear and are given
by the Szekeres~\cite{Szekeres} family, including the subcase of
the Ellis~\cite{Ellis} LRS class II dust models and such well
known examples as the Lemaitre-Tolman-Bondi model and the
orthogonally spatially homogeneous Kantowski-Sachs model. These
models can be easily generalized to take into account a non-zero
cosmological constant \cite{Barrow, Szafron1, Szafron2}. \\

For the general Petrov type I case the situation is rather
different, the only known silent universe being the orthogonally
spatially homogeneous Bianchi I model. The issue of the
integrability conditions of the inhomogeneous silent universes of
Petrov type I has been a source of confusion and discussion during
the last decade~\cite{Bruni, Lesame, Sopuerta, vanElst, Mars} and,
although the field equations of the silent models are consistent
when linearised about a Friedman-Robertson-Walker background, the
full non-linear consistency problem remains --- to this day ---
unsolved. A detailed study of the constraints so far only has led
to the \emph{conjecture} that the Bianchi type I models are the
unique Petrov type I silent universes.\\

At first sight it looks as if the presence of a cosmological
constant is of little relevance in the consistency problem (it
might well be that this is indeed so in the \emph{generic}
situation). However, one of the special cases where one can show
that the constraints with $\Lambda=0$ are inconsistent (namely the
case where one of the eigenvalues of the expansion tensor is 0),
surprisingly turns out to admit explicit type I solutions when
$\Lambda > 0$.

\section{Basic equations}

We first interpret the cosmological constant as a constant
negative pressure $p=-\Lambda$ and rewrite the field equations
(\ref{veldvergelijkingen}) in the form (also substituting $\mu$ by
$\mu +p\neq0$)
\begin{equation}\label{veldvergelijkingen2}
R_{ab}-\frac{1}{2}R
g_{ab}= (\mu+p) u_a u_b +p g_{ab}
\end{equation}
and we further follow the notations and conventions of the
Ellis-MacCallum orthonormal tetrad formalism~\cite{MacCallum}.
Herein, the normalized 4-velocity $u^a$ plays the role of the
timelike basis vectorfield $e_0^a$ of the tetrad
$\{e_0^a,\mathbf{e}_\alpha^a\}$. Greek indices take the values 1,
2 and 3 and refer to tetrad components with respect to
$\mathbf{e}^a_{\alpha}$. Consequently, for spatial tensorfields
(i.e.\ for which all contractions with $u^a$ vanish) only the
Greek indexed components are not a priori zero. For spatial
vectorfields $v_a$ and rank 2 tensorfields $S_{ab}$ these
components $v_\a$ and $S_{\a\b}$ form $(3\times 1)$ resp.\
$(3\times 3)$ matrices denoted by boldface symbols $\mathbf{v}$
and $\mathbf{S}$ and we will write $S_\a$ for the eigenvalues of
$\mathbf{S}$. As a trivial example, the tensor $h_{ab}=g_{ab}+u_a
u_b$, which projects orthogonally to $u^a$, satisfies
$\mathbf{h}=\mathbf{1}$, i.e.\ $h_{\a\b}=\delta_{\a\b}$.

The basic variables in this formalism are 24 independent linear
combinations of the Ricci rotation coefficients or, equivalently,
of the commutator coefficients $\gamma^a_{bd}$ associated with the
tetrad; these are the objects $n_{\a\b}$ ($\a \leq \b$, with
additionally $n_{\b\a}=n_{\a\b}$) and $a_\a$ defined by
$\gamma^\a_{\b\d}=\epsilon_{\b\d\e}n^{\e\a}+\delta^\a_\d a_\b -
\delta^\a_\b a_\d$, the components $\Omega_\a$ of the angular
velocity of the triad $\mathbf{e}_\a$ with respect to the local
`inertial compass' and the components $\dot{u}_\a$, $\omega_{a}$
and $\theta_{\a\b}$ of the spatial kinematic quantities
(acceleration vector, vorticity vector and expansion tensor
respectively). We also use the decomposition
$\theta_{\a\b}=\sigma_{\a\b}+1/3 \Theta\delta_{\a\b}$, with
$\sigma_{ab}$ the trace-free shear tensor and $\Theta$ the
expansion scalar. It turns out that there is a slight notational
advantage in using instead of the variables $n_{\alpha \beta}\
(\alpha\neq\beta)$ and $a_\alpha$ the variables $q_\alpha$ and
$r_\alpha$ defined by
\begin{equation*}
n_{\alpha +1\ \alpha-1}=2 q_\alpha+r_\alpha, \
a_\alpha=r_\alpha-q_\alpha,
\end{equation*}
Here and below these expressions have to be read modulo 3, so for
example $\alpha = 3$ gives $n_{12}=2 q_3 + r_3$.\\
The basic equations in the formalism are the Jacobi-equations
(77)-(81) and the field equations (82)-(84) of~\cite{MacCallum}
(where the print error in (83) was corrected
in~\cite{Vandenbergh}) together with the energy-momentum
conservation equations or 'contracted' Bianchi identities. Below,
the labels (77)-(84) refer to the corresponding equations
 in~\cite{MacCallum}.\\

For problems in which conditions on $E$ and $H$ are imposed one
can proceed as follows. Both tensorfields being spatial, the only
not a priori vanishing components are
\begin{equation}
E_{\a\b} = C_{\a 0\b 0} \ \textrm{ and } H_{\a\b} = \frac{1}{2}
{\eta _{\a 0}}^{mn} C_{mn\b 0}
\end{equation}
and can be expressed in terms of the basic variables and their
derivatives in two ways. On the one hand, one can take the
definition of the Weyl tensor through the decomposition of the
curvature tensor (see (3.45) or (3.50) of \cite{Kramer}), using
the formula for the curvature components in terms of the Ricci
rotation coefficients and their derivatives (see the appendix) and
using the expressions $R_{\a\b}$ for the Ricci tensor components
originating from the field equations (\ref{veldvergelijkingen2}),
to yield expressions which we denote by $E^C_{\a\b},\ H^C_{\a\b}$;
on the other hand one can deduce~\cite{Ellis} expressions
$E^u_{\a\b},\ H^u_{\a\b}$ from the Ricci identities for $u^a$,
which are given in $1+3$ covariant form by (A3) and (A6)
of~\cite{Maartens}. Both ways are of course equivalent via the
Einstein and Jacobi equations: one can check that
\begin{equation}\label{verschilEenH}
\fl \eqalign{
E^u_{\a\a}-E^C_{\a\a} =0 & \Longleftrightarrow \frac{1}{3}(82),\\
E^u_{\a +1\, \a -1}-E^C_{\a +1\, \a -1} =0 & \Longleftrightarrow \frac{1}{2}(79)^\a ,\\
 E^u_{\a -1\, \a +1}-E^C_{\a -1\, \a +1}=0 &  \Longleftrightarrow -\frac{1}{2}(79)^\a,\\
H^u_{\a\a}-H^C_{\a\a}=0 & \Longleftrightarrow \frac{1}{2}[(81)^{\a\a} -
(81)^{\a +1\, \a +1} -
(81)^{\a -1\, \a -1}] + \frac{1}{6} (77),\\
H^u_{\a +1\, \a -1}-H^C_{\a +1\, \a -1}=0 & \Longleftrightarrow
\frac{1}{2}(80)^\a + (81)^{\a +1\, \a -1} - \frac{1}{2}(83)^\a \\
H^u_{\a -1\, \a +1}-H^C_{\a-1\, \a +1} & \Longleftrightarrow
-\frac{1}{2}(80)^\a + (81)^{\a +1\, \a -1} +\frac{1}{2}(83)^\a,}
\end{equation}
where for example $(80)^3$ and $(81)^{12}$ mean `$(80)$ with $\g=3$' and `$(81)$ with
$\a=1,\b=2$' respectively.

These components are treated as extra variables and form matrices
$\mathbf{E}$ and $\mathbf{H}$, the symmetry and tracelessness of
which is automatically fulfilled when using $\mathbf{E}^u$ and
$\mathbf{H}^u$, but for $\mathbf{E}^C$ and $\mathbf{H}^C$ only
follows from specific Einstein and Jacobi equations --- as is
evident from (\ref{verschilEenH}).

Finally, the 16 remaining Bianchi identities~\cite{Ellis} are added to the basic equations;
they can most easily be translated in terms of the 
above variables from their 1+3 covariant forms (e.g. (A7)-(A12)
of~\cite{Maartens}) by means of the connecting formulas presented
in the appendix. Below, the labels (Ax) refer to the corresponding
equations in~\cite{Maartens}.

Returning to the case of silent universes, we have to cope with
the basic equations under the additional conditions
\begin{equation}\label{silentcondities}
\mathbf{H}=0, \quad \mathbf{\omega}=0, \quad \dot{\mathbf{u}}=0,
\quad\Theta\neq 0,\quad (E_1-E_2)(E_2-E_3)(E_3-E_1)\neq 0,
\end{equation}
the last condition expressing the fact that we restrict ourselves
to the Petrov type I case; the third condition follows directly
from the energy-momentum conservation equation (A8), while (A7)
determines the evolution of the matter density $\mu$ along the
matter flow lines,
\begin{equation}\label{d0mu}
\dot{\mu}\equiv \partial_0 \mu =-(\mu + p)\Theta .
\end{equation}

We next recall some basic results from \cite{Barnes}. Under the
conditions (\ref{silentcondities}) the $\textrm{div}\mathbf{H}$
Bianchi identities (A12) become algebraic,
\begin{equation*}
\epsilon_{\alpha\beta\gamma}\sigma^{\beta\delta}{E_{\delta}}^\gamma=0
\end{equation*}
($\epsilon_{\a\b\g}$ denotes the usual 3-dimensional Levi-Civita
skew symbol) and expresses exactly that the antisymmetric part of
$\mathbf{\sigma}\cdot \mathbf{E}$ vanishes, i.e.\ that
$\mathbf{\sigma}$ and $\mathbf{E}$ commute. Therefore it becomes
advantageous to work in an orthonormal eigenframe of $\mathbf{E}$,
which thus is simultaneously an eigenframe of $\mathbf{\sigma}$
and $\mathbf{\theta}$: this fixes the triad $\{\mathbf{e}_\alpha\}$
completely since the Petrov type is I and drastically decreases
the number of variables. First, the sets $\{E_{\a\b}\}$ and
$\{\theta_{\a\b}\}$ are reduced to the respective eigenvalue sets
$\{E_{\a}\}$ and $\{\theta_{\a}\}$. Second, the off-diagonal part
of (A9)
\begin{equation}
\dot{E}_{\a\b}\equiv
\partial_0 E_{\a\b}+2\Omega^{\gamma}\epsilon_{\gamma\delta(\alpha}{E_{\beta)}}^\delta =-\Theta
E_{\a\b}+3{\sigma^\gamma}_{\langle\a}E_{\b\rangle\g}
-\frac{1}{2}(\mu+p)\sigma_{\a\b}
\end{equation}
shows that an eigenframe of $\mathbf{E}$ and $\mathbf{\sigma}$ is
also Fermi-propagated along the matter flowlines, i.e.
$\mathbf{\Omega}=0$. Third, as shown in~\cite{Barnes} for Petrov
type I fields, the diagonal parts of (A6) and the
$\mathbf{\dot{H}}$ Bianchi identities (A10),
\begin{equation}\label{HenHpunt}
0=H^u_{\a\a}=(\textrm{curl}\sigma)_{\alpha\alpha},\quad
0=\dot{H}^u_{\alpha\alpha}=-(\textrm{curl} E)_{\alpha\alpha}
\end{equation}
(where curl is the covariant spatial curl operator, the definition
of which is also given in~\cite{Maartens}), are equivalent with
the fact that the Weyl principal vectors are hypersurface
orthogonal, i.e\ with the vanishing of all Ricci rotation
coefficients $\gamma_{\alpha-1 \ \alpha+1}^\alpha$ or, in our
formalism, with $n_{\alpha \alpha}=0$. More precisely, Barnes
shows that (\ref{HenHpunt}) implies either this fact \emph{or}
else that the fluid is shear free ($\mathbf{\sigma}=0$) and
thus~\cite{Barnes2} static, the latter possibility being
contradictory to the condition $\Theta\neq 0$).

Under the previous conditions the Jacobi equations (77), (79) and the diagonal part of (81)
are identically fulfilled, so that the
the diagonal components of $\mathbf{H}^u$ and $\mathbf{H}^C$ are
the same (see (\ref{verschilEenH})) and vanish identically. There
are neither extra calculations involved in expressing
$H^C_{\a\b}=0$, which implies the evolution equations

\begin{equation}
\eqalign{
\partial_0 r_\alpha &= -\theta_{\alpha+1}r_\alpha \\
\partial_0 q_\alpha &= -\theta_{\alpha-1}q_\alpha .
} \label{d0renq}
\end{equation}

The $E^C_1$ expression for $E_1$ together with the (0 0) field
equation (82) give

\begin{equation}\eqalign{
\partial_0 \theta_\alpha & = - E_\alpha
-\theta_\alpha^2-\frac{1}{6}(\mu+3 p) \label{H_ev} \\
E_{33} & = -(E_{11}+E_{22}). }\label{d0theta}
\end{equation}
Substituting the expressions (\ref{d0renq}) in the Jacobi
equations $(80)^\a$ and $(81)^{\a +1\a -1}$, these readily
determine the spatial derivatives $\partial_\alpha
\theta_{\alpha+1}$ and $\partial_\alpha \theta_{\alpha-1}$ to be
\begin{equation}
\eqalign{
\partial_\alpha \theta_{\alpha+1} & = 2 r_\alpha
(\theta_{\alpha+1}-\theta_{\alpha}) \\
\partial_\alpha \theta_{\alpha-1} & = 2 q_\alpha
(\theta_{\alpha}-\theta_{\alpha-1}). } \label{partialstheta}
\end{equation}
Under (\ref{d0renq}) and (\ref{partialstheta}) the
$(0\ \alpha)$ field equations (83) become identities,
see (\ref{verschilEenH}). The $(\a +1\ \a -1)$  field equations together with the
only remaining Jacobi equations $(78)^\a$ imply

\begin{equation}
\eqalign{
\partial_{\alpha} r_{\alpha+1} & = -2 q_\alpha
(q_{\alpha+1}+ r_{\alpha+1}) \\
\partial_\alpha q_{\alpha-1} & = 2 r_\alpha
(q_{\alpha-1}+ r_{\alpha-1})}, \label{partialsrenq}
\end{equation}

while, with (\ref{d0theta}), the $(\alpha \alpha)$ field equations
yield
\begin{equation}\label{einaa}
\partial_\alpha r_\alpha- \partial_{\alpha+1}
q_{\alpha+1}= 2 r_\alpha^2+2 q_{\alpha+1}^2-2
r_{\alpha-1}q_{\alpha-1}
-\frac{1}{2}\theta_\alpha\theta_{\alpha+1}-\frac{1}{2}E_{\alpha-1}+\frac{1}{6}
\mu. \end{equation}

At this point all Jacobi and Einstein equations have been dealt with. Finally, the diagonal
part of the $\dot{\mathbf{E}}$ Bianchi identities (A9) yields
\begin{equation}\label{d0E}
\partial_0 E_\alpha = (\theta_\alpha-\theta) E_\alpha
+E_{\alpha+1} \theta_{\alpha-1} + E_{\alpha-1}
\theta_{\alpha+1}+\frac{1}{6}(\mu+p)(\theta-3 \theta_\alpha),
\end{equation}
while the $\textrm{div}\mathbf{E}$ Bianchi
identities (A11), and then the off-diagonal part of (A10), lead to the following expressions,
where the variables $m_\alpha$ represent the spatial gradients of
the matter density, $m_\alpha=\partial_\alpha \mu$:
\begin{equation}
\eqalign{
\partial_\alpha E_\alpha &= \frac{1}{3}m_\alpha-2 r_\alpha
(E_{\alpha+1}-E_\alpha)-2 q_{\alpha} (E_\alpha-E_{\alpha-1}) \\
\partial_\alpha E_{\alpha+1} &= -\frac{1}{6} m_\alpha +2 r_\alpha
(E_{\alpha+1}-E_\alpha) \\
\partial_\alpha E_{\alpha-1} &= -\frac{1}{6} m_\alpha -2q_\alpha
(E_{\alpha-1}-E_\alpha)\,.}
\end{equation}

Herewith all basic equations are satisfied identically.

\section{Integrability conditions}
Propagating the expressions for the spatial gradients of
$E_\alpha$ along the matter flow lines, and using the commutator relations (75)
of ~\cite{MacCallum} and the above deduced equations, one obtains
\begin{equation}
\fl (E_{\alpha-1}-E_{\alpha+1}) \partial_\alpha \theta_\alpha= 6
E_\alpha [q_\alpha (\theta_{\alpha-1}-\theta_\alpha)+r_\alpha
(\theta_{\alpha+1}-\theta_\alpha)]
+\frac{1}{2}m_\alpha (\theta_{\alpha+1}-\theta_{\alpha-1})
\end{equation}
and
\begin{eqnarray}
\fl (E_{\alpha-1}-E_{\alpha+1})\partial_0 m_\alpha =4 (\mu+p)[
q_\alpha (E_\alpha-E_{\alpha-1})(\theta_\alpha-\theta_{\alpha-1})
+ r_\alpha
(E_\alpha-E_{\alpha+1})(\theta_\alpha-\theta_{\alpha+1})]\nonumber \\
+\frac{1}{2}m_\alpha (\mu+p) (\theta_{\alpha-1}-\theta_{\alpha+1})
-m_\alpha (\theta+\theta_\alpha)(E_{\alpha-1}-E_{\alpha+1})
\end{eqnarray}
Further integrability conditions can be obtained by considering
the $[\partial_{\alpha-1},\ \partial_{\alpha+1}] E_\alpha$
commutator
relations, but this is at present not our concern.\\

As Barnes noticed in~\cite{Barnes}, the expansion tensor of type I
silent universes is non-degenerate: the time evolution of, for
example, $\theta_1=\theta_2$ would immediately imply $E_1=E_2$,
see (\ref{d0theta}). Looking at the equations
(\ref{d0mu}),(\ref{d0theta}) and (\ref{d0E}) one sees that
$\partial_0$ is a derivation on the polynomial algebra $\mathbb{R}
[E_{1},E_{2},\theta_{1},\theta_{2},\theta_3, \mu]$. Therefore it
is fairly straightforward to investigate whether a certain
algebraic relation between the 6 mentioned variables is
consistent: to obtain new algebraic relations one can take time
derivatives of a newly generated algebraic relation or calculate
the resultant of it and an already deduced algebraic relation
(with respect to an appropriate variable which thereby is
eliminated from these two). From the above, the most natural
\emph{Ansatze} seem to be $\theta_1=0$ or $E_1=0$. However, for
the last one we get the same inconsistency result as for general
vacua (cf.\ \cite{Brans}):\\

\textbf{Proposition} There exist no silent universes (with or
without cosmological constant) for which one of the eigenvalues of
the Weyl tensor vanishes.\\

\textbf{Sketch of proof:} We denote the resultant of two
polynomial expressions $P_1[x_1,\ldots,x_n]$ and
$P_2[x_1,\ldots,x_n]$ with respect to the variable $x_i$ by
$\textrm{res}_{x_i}(P_1,P_2)[x_1, \ldots,\hat{x_i},\ldots,x_n]$
(where $\hat{x_i}$ means that the variable $x_i$ is supprimed).
Suppose e.g.\ $E_{1}=0$. From (\ref{d0E}) with $\a=1$ we get
\begin{equation}
P_1[E_{2},\theta_{1},\theta_{2},\theta_3,\mu]:=\frac{\mu +
p}{6}(-2 \theta_{1}+\theta_{2}+ \theta_{3})+(\theta_{3}-\theta_{2})E_{2}=0.
\end{equation}
Put $P_2[\theta_{1},\theta_{2},\theta_3,\mu]:=18
\textrm{res}_{E_2}(\partial_0 P_1,P_1)
[\theta_{1},\theta_{2},\theta_3,\mu]/(\mu +p)$ and
$P_3[\theta_{1},\theta_{2},\theta_3,\mu]:=
\textrm{res}_{E_2}(\partial_0
P_2,P_1)[\theta_{1},\theta_{2},\theta_3,\mu]/(\theta_2 -
\theta_3)$. Herewith the resultant
$\textrm{res}_{\mu}(P_2,P_3)[\theta_{1},\theta_{2},\theta_3]$
yields
\begin{equation}
  3(\theta_2 - \theta_3)^4(2 \theta_1-\theta_2-\theta_3)
   (6 \theta_1^2-3 \theta_2^2-3 \theta_3^2-7 \theta_1 \theta_2-7
\theta_1  \theta_3+14 \theta_2 \theta_3) =0.
\end{equation}
The first possibility to satisfy this relation,
$\theta_3=\theta_2$, is not allowed because of the non-degeneracy
of the expansion tensor; the second possibility
$2\theta_1-\theta_2-\theta_3= 0$ comes down to the first one,
since expressing $\partial_0(2 \theta_1-\theta_2-\theta_3)$ and
substituting $\theta_1=(\theta_2 + \theta_3)/2$ yields
$-(\theta_2-\theta_3)^2 /2$, from which we get
\begin{equation}
P_4[\theta_{1},\theta_{2},\theta_3]:=6 \theta_1^2-3 \theta_2^2-3
\theta_3^2-7 \theta_1 \theta_2-7 \theta_1  \theta_3+14 \theta_2
\theta_3=0.
\end{equation}
Here, $P_4$ is homogeneous and quadratic in
$\theta_{1},\theta_{2}$ and $\theta_3$, but not positive definite.
Therefore we proceed and calculate successively \begin{equation}
\eqalign{\textrm{res}_{E_2} (\partial_0
P_4,P_1)[\theta_{1},\theta_{2},\theta_3,\mu ] &
:=(\theta_2 - \theta_3)Q_1[\theta_{1}, \theta_{2},\theta_3,\mu]/6\\
\textrm{res}_{\mu}(Q_1,P_2)[\theta_{1},\theta_{2},\theta_3 ] & :=
(2\theta_1-\theta_2-\theta_3)Q_2[\theta_{1},\theta_{2},\theta_3]}\end{equation}

and get
\begin{equation}
\textrm{res}_{\theta_3}(Q_2,P_4)[\theta_{1},\theta_{2},\theta_3):=-96(\theta_{1}-\theta_{2})^4
P_5[\theta_1,\theta_2],
\end{equation}
where $P_5$ is a polynomial of total degree 4 in $\theta_1$ and
$\theta_2$. Proceeding in exactly the same way for $P_5$ as for
$P_4$, one gets polynomials
$R_1[\theta_{1},\theta_{2},\theta_3,\mu]$ and
$R_2[\theta_{1},\theta_{2},\theta_3]$. Finally, elimination of
$\theta_3$ and $\theta_2$ from $R_2$, making use of $P_4$ and
$P_5$ respectively, results in a polynomial of degree 40 in
$\theta_1$, with coefficients depending on $p=-\Lambda$. This
implies $\theta_1$ is constant and hence, by $P_5$, also
$\theta_2$ is constant. But then $\mu$ is constant by
(\ref{d0theta}) (for $\a = 1$) and thus $\Theta=0$ from
(\ref{d0mu}), which is the desired contradiction.\\


Most likely, extended classes of the previous \emph{Ansatz} (e.g.\
$E_1=a E_2$, $a$ constant) can be proven to be inconsistent by the
same technique, but we do not consider these here. \\

For the other case $\theta_1=0$ we can start as above, but now we
get a qualitatively different picture.

From $\partial_0 \theta_1=0$ one obtains
\begin{equation}
E_{1}=-\frac{1}{6}(\mu+3 p),
\end{equation}
two further time
derivatives of which yield consecutively
\begin{equation}
E_2=\frac{(\mu+3 p)(\theta_3+2 \theta_2)}{6(\theta_2-\theta_3)}
\end{equation}
and
\begin{equation}
(\mu+3p)^2+4 p (\theta_2-\theta_3)^2=0. \label{matter}
\end{equation}
For $p\geqslant 0$ (i.e. the case of vanishing or negative
cosmological constant), this leads to $\mu + 3p=0$ and thus to the
inconsistency $E_1=E_2=0$. Taking a further time derivative of
(\ref{matter}), shows that it is conserved along the matter flow
lines, suggesting to have a closer look at the possible existence
of solutions for $p<0$, i.e. $\Lambda > 0$.

\section{Solutions with  $\theta_1=0$ and $\Lambda > 0$}
With $\theta_1=0$ one deduces from (\ref{partialstheta}) that
$q_2=r_3=0$. Herewith $\partial_1 \theta_1=0$ gives an algebraic
expression for $m_1$, which turns out to be conserved along the
matter flowlines under the constraint (\ref{matter}):
\begin{equation}
 m_1= 2(\mu+3 p)\frac{\theta_3 q_1+\theta_2
 r_1}{\theta_2-\theta_3}.
\end{equation}
Whereas no new information enters by propagating the constraint
(\ref{matter}) along the $\partial_1$ integral curves, propagation
along $\partial_2$ and $\partial_3$ shows that also $q_3=r_2=0$.
Herewith the expressions (\ref{einaa}) for $\a=2$ and $\a=3$ result in
\begin{equation}
\partial_1 q_1 = -2 q_1^2 +\frac{1}{2} p +\frac{\mu+3 p}{4} \frac{\theta_3}{\theta_2-\theta_3}
\end{equation}
and
\begin{equation}
\mu + p - 2 \theta_2 \theta_3 - 8 q_1 r_1=0. \label{qr}
\end{equation}
It is advantageous now to introduce a constant $\lambda$
($=\sqrt{\Lambda}$), defined by $p=-\lambda^2$ and allowing one to
rewrite (\ref{matter}) as

\begin{equation}
\mu = 3 \lambda^2 +2 \lambda(\theta_2-\theta_3).
\end{equation}
Herewith (\ref{qr}) becomes
\begin{equation}
4 q_1 r_1+(\theta_2+\lambda)(\theta_3-\lambda) = 0\label{qr1}
\end{equation}

and the remaining equations for the variables $q_1, r_1, \theta_2,
\theta_3$ and $\mu$ simplify to

\begin{equation}\label{vgl_qr}
\left\{\eqalign{
\partial_0 q_1 &=  -\theta_3 q_1 \\
\partial_1 q_1 &=  -2 q_1^2-\frac{1}{2} \lambda(\lambda- \theta_3)\\
\partial_2 q_1 &=  0}\right.
\ \ \
\left\{\eqalign{
\partial_0 r_1 &=  -\theta_2 r_1 \\
\partial_1 r_1 &=  2 r_1^2+ \frac{1}{2} \lambda(\lambda+ \theta_2)\\
\partial_3 r_1 &=  0}\right.
\end{equation}

\begin{equation}\label{vgl_H23}
\left\{\eqalign{
\partial_0 \theta_2 &=  -\theta_2(\lambda+\theta_2)\\
\partial_1 \theta_2 &=  2 \theta_2 r_1\\
\partial_3 \theta_2 &=  0}\right.
\ \ \
\left\{\eqalign{
\partial_0 \theta_3 &=  \theta_3(\lambda-\theta_3)\\
\partial_1 \theta_3 &=  -2 \theta_3 q_1\\
\partial_2 \theta_3 &=  0.}\right.
\end{equation}

It is clear that $q_1=0$ iff $\theta_3=\lambda$, while $r_1=0$ iff
$\theta_2=-\lambda$. It is easy to check that the integrability
conditions for the remaining variables are then identically
satisfied.\\
When $q_1 r_1 \neq 0$ the integrability conditions for the
equations (\ref{vgl_qr}, \ref{vgl_H23}) imply further relations,
which allow one to close the system by introducing new variables
$u=\theta_2^{-3}\partial_2 \theta_2$ and
$v=\theta_3^{-3}\partial_3 \theta_3$, together with their higher
order derivatives $u \theta_2 u'=\partial_2 u, v \theta_3
v'=\partial_3 v$ and ${u'}^{2} \theta_2 u''=\partial_2 u',
{v'}^{2} \theta_3 v''=\partial_3 v'$. The following extra
equations can then be generated:

\begin{equation}
\partial_3 q_1 = -\frac{(\theta_2+\lambda)v}{4 r_1 }\theta_3^3,\ \ \
\partial_2 r_1 = -\frac{(\theta_3-\lambda)u}{4 q_1 }\theta_2^3
\end{equation}

\begin{equation}\label{vgl_uv}
\left\{\eqalign{
\partial_0 u &= 2 \lambda u\\
\partial_1 u &= -2 \lambda u \frac{r_1}{\lambda+\theta_2}\\
\partial_3 u &= 0}
\right.
\ \ \
\left\{ \eqalign{
\partial_0 v &= -2 \lambda v\\
\partial_1 v &= 2 \lambda v \frac{q_1}{\lambda-\theta_3}\\
\partial_2 v &= 0}
\right.
\end{equation}

\begin{equation}\label{vgl_u'}
\left\{\eqalign{
\partial_0 u' &= \lambda u'\\
\partial_1 u' &= 0\\
\partial_3 u' &= 0}\right.
\ \ \
\left\{ \eqalign{
\partial_0 v' &= - \lambda v'\\
\partial_1 v' &= 0\\
\partial_2 v' &= 0}
\right.
\end{equation}

and

\begin{equation}
\partial_0 u''= \partial_1 u'' =\partial_3 u'' = \partial_0
v''=\partial_1 v'' =\partial_3 v''=0.
\end{equation}

The complete set of integrability conditions for these equations
implies ---and is identically satisfied under--- the relations
\begin{equation}\partial_2 u'' = \theta_2 u' \Phi(u''),\ \ \partial_3 v'' =
\theta_3 v' \Psi(v'')\end{equation} with $\Phi$ and $\Psi$
arbitrary functions.\\

Notice that the eigenvalues of $\mathbf{E}$ are given by

\begin{equation}
E_1=-\frac{\lambda}{3}(\theta_2-\theta_3),
E_2=\frac{\lambda}{3}(\theta_3+2\theta_2),
E_3=-\frac{\lambda}{3}(\theta_2+2\theta_3),
\end{equation}

clearly showing that the Petrov type is I (unless $\theta_2$ or
$\theta_3=0$; $\theta_2+\theta_3=0$ is excluded as this would
imply $\Theta=0$).\\

One can easily show (using the fact that the frame is fixed and
all variables are invariants) that in general, with $r_1 q_1\neq
0\neq u v$, the corresponding solutions will only admit a group of
isometries when $\Phi=\Psi=\frac{1}{2}$. In that case it follows
that
\begin{equation}
u'=-2\lambda u\frac{\theta_2}{\theta_2+\lambda}, v'=2\lambda v
\frac{\theta_3}{\theta_3-\lambda}
\end{equation}
and the ---in general unique--- Killing vector is given by
\begin{equation}
\mathbf{K}=\partial_0+\frac{\theta_2+\lambda}{u\theta_2^2}
\partial_2+\frac{\theta_3-\lambda}{v \theta_3^2} \partial_3.
\end{equation}
When $q_1 r_1=0$ or $u v =0$ an extra Killing vector may arise (cf. below).\\

Using the fact that the tetrad basisvectors are
hypersurface-orthogonal~\cite{Barnes}, it is straightforward to
construct the
metric\footnote{A coordinate transformation is used to reduce the coefficients of $e^{\pm \lambda t}$
to 1; when these coefficients are 0 one obtains the Petrov type D subcase with
$\theta_1=\theta_2=0$ or $\theta_1=\theta_3=0$.}: \\

when $q_1 r_1 \neq 0$ one finds, with $f_1(y)$ and $g_1(z)\neq 0$,

\begin{equation}\label{generalmetric}
\fl ds^2=-dt^2+dx^2+ (e^{-\lambda t} + f_1(y)\cos \lambda x)^2
dy^2+ (e^{\lambda t} + g_1(z)\sin \lambda x)^2 dz^2,
\end{equation}
with matter density given by

\begin{equation}
\fl \mu= \lambda^2 \frac{3 f_1(y)g_1(z)\sin \lambda x \cos \lambda
x+e^{\lambda t} f_1(y) \cos \lambda x+e^{-\lambda t} g_1(z)\sin
\lambda x -1}{(e^{-\lambda t}+ f_1(y) \cos \lambda x)(e^{\lambda
t}+g_1(z)\sin \lambda x)},
\end{equation}

whereas for $q_1=0$ or $r_1=0$ one obtains
\begin{equation}\label{specialmetric}
\fl ds^2=-dt^2+dx^2+ [e^{-\lambda t} + f_1(y)\cos (f_2(y)+\lambda
x)]^2 dy^2+ e^{2 \lambda t}dz^2.
\end{equation}

with matter density given by \begin{equation} \fl \mu = \lambda^2
\frac{f_1(y)\cos (f_2(y)+\lambda x)+e^{-\lambda t}}{f_1(y)\cos
(f_2(y)+\lambda x)-e^{-\lambda t}}.
\end{equation}

The solution (\ref{generalmetric}) admits a (unique) Killing
vector when either $f_1$ or $g_1$ is a constant \emph{or} when
$f_1(y)\sim y^{-1}$ and $g_1(z)\sim z^{-1}$. It admits a
two-dimensional group of isometries when both $f_1$ and $g_1$ are
constants. The solution (\ref{specialmetric}) obviously admits a
Killing vector, aligned with $\mathbf{e}_2$ or $\mathbf{e}_3$ and
will admit a second Killing vector when $f_1$ and $f_2$ are
constants.

\section{Conclusion}

We have shown that, in contrast to the case of zero cosmological
constant, where Petrov type I silent dust universes appear to be
restricted to the orthogonally spatially homogeneous Bianchi type
I , for positive cosmological constant 'large' classes of
solutions exist (in the sense that they contain arbitrary
functions of the coordinates). The key equation in the
construction is equation (\ref{matter}), which relates the matter
density to the eigenvalues of the expansion tensor and follows
from imposing the \emph{Ansatz} that one of the eigenvalues of the
latter is 0. It clearly shows also why such an \emph{Ansatz} does
not lead to
solutions when $\Lambda =0$.\\

The solutions could be explicitly obtained and are given by the
solutions (\ref{generalmetric}) and (\ref{specialmetric}). It is
an open problem whether these solutions are the only $\Lambda \neq
0$ ones of Petrov type I.

\section{Appendix}

In order to avoid confusion with different conventions for the
Ricci rotation coefficients, we prefer to define the kinematical
quantities via the connection 1-forms ${\mathbf{\Gamma}^a}_{b}$.
For the tetrad $\{e_a\}$ and dual basis $\{e^a\}$ the latter are
defined by
\begin{equation}
{\mathbf{\Gamma}^a}_{b}={\Gamma^a}_{bc}e^c=({e^a}_j {{e_b}^j}_{;i}
{e_c}^i)e^c.
\end{equation}
With $\mathbf{\Gamma}_{ab}=g_{ac} {\mathbf{\Gamma}^c}_{b}$ (such
that $\mathbf{\Gamma}_{ba}=-\mathbf{\Gamma}_{ab}$) and with
$u^i=e_0^i$, the components of these 1-forms with respect to
$\{e^a\}$ are then given in terms of the variables appearing
in~\cite{MacCallum} by
\begin{equation}
\eqalign{ {\Gamma}_{\alpha 0 0} =\dot{u}_\alpha, & \
{\Gamma}_{\alpha 0\beta}=\theta_{\a\b}+\epsilon_{\a\b\g}\omega^\g \\
{\Gamma}_{\a\b 0}= \epsilon_{\b\a\g}\Omega^\g, & \
{\Gamma}_{\a\b\d}=1/2\epsilon_{\a\b\g}{n^{\g}}_\d-{\epsilon^\g}_{\d[\a}n_{\b]\g}-
2\delta_{\d[\a}a_{\b]}.}
\end{equation}
From this, the expression
\begin{equation}
R_{abcd}=\partial_c \Gamma_{abd}-\partial_d {\Gamma}_{abc}+
{\Gamma}_{aec} {{\Gamma}^e}_{bd}-{\Gamma}_{aed}{{\Gamma}^e}_{bc}+
{\Gamma}_{abe}({{\Gamma}^e}_{cd}-{{\Gamma}^e}_{dc})
\end{equation}
and the decomposition (3.50) in~\cite{Kramer}, one can derive
expressions for the components of the Weyl tensor (and thus of its
electric and magnetic part) in these variables, as mentioned in
the text.

For general purpose, we give also the following connecting
formulas (the covariant spatial derivative $D$, divergence
operator div and curl operator curl appearing herein are defined
in e.g. \cite{Maartens} and are acting on a spatial vector $v^a$
and a spatial and symmetric rank 2 tensor $S_{ab}$):
\begin{equation*} \fl \eqalign{
\textrm{D}_{\alpha} v_{\beta}  = \partial_\alpha v_\beta -
1/2\epsilon_{\alpha\beta\gamma} n^{\gamma\delta}v_\delta +
v_\gamma{\epsilon^{\gamma\delta}}_{(\alpha} n_{\beta)\delta} +
v_\alpha a_\beta - \delta_{\alpha\beta}a_\gamma v^\gamma\\
\textrm{div}\, v =  \partial_\gamma v^\gamma - 2 a_\gamma v^\gamma\\
\textrm{curl}\, v_\alpha =
\epsilon_{\alpha\beta\gamma}(\partial^\beta v^\gamma-a^\beta
v^\gamma) - n_{\alpha\gamma} v^\gamma\\
\dot{v}_\a = \partial_0 v_\a+\epsilon_{\a\b\g}\Omega^\beta v^\gamma \\
\textrm{div}\, S_\alpha =
\partial_\beta {S_\alpha}^\beta-3a_\beta {S_\alpha}^\beta-
\epsilon_{\alpha\beta\gamma}n^{\beta\delta}{S_\delta}^\gamma + \textrm{tr}S\, a_\alpha\\
\textrm{curl}\, S_{\alpha\beta} = (\partial_\gamma
S_{\delta(\alpha}-a_\gamma
S_{\delta(\alpha}){\epsilon_{\beta)}}^{\gamma\delta} - 3
n_{\gamma\langle\alpha} {S_{\beta\rangle}}^\gamma + \textrm{tr}S\,
(n_{\alpha\beta}-1/2 {n_\gamma}^\gamma\delta_{\alpha\beta})+1/2
{n_\gamma}^\gamma S_{\alpha\beta}\\
\dot{S}_{\a\b} = \partial_0 S_{\a\b}+2\epsilon_{\g\d(\a}\Omega^\g
{S_{\b)}}^\d.}
\end{equation*}

\end{document}